\documentclass[12pt]{article}
\usepackage{epsfig}
\usepackage{float}
\floatname{figure}{Figure}
\newcommand{\nc}{\newcommand}
\def\frac#1#2{{\textstyle {#1 \over #2}}}

\nc{\beq}{\begin{equation}}
\nc{\eeq}{\end{equation}}
\nc{\beqa}{\begin{eqnarray}}
\nc{\eeqa}{\end{eqnarray}}
\nc{\lsim}{\begin{array}{c}\,\sim\vspace{-21pt}\\< \end{array}}
\nc{\gsim}{\begin{array}{c}\sim\vspace{-21pt}\\> \end{array}}
\nc{\appA}{}
\nc{\appB}{}
\nc{\appC}{}
\nc{\appD}{}
\nc{\appE}{}
\nc{\Eqr}[1]{(\ref{#1})}
\nc{\mysection}[1]{\setcounter{equation}{0}\section{#1}}

\nc{\myappendix}[1]{\section*{#1}\setcounter{equation}{0}}

\def\bra{\langle }
\def\ket{\rangle }

\def\dk{d\bar{k}}
\def\dr{d\bar{r}}
\def\dq{d\bar{q}}

\def\bpsi{\bar\psi}

\def\bta{\bar\eta}

\def\&{and}

\def\kS{k\!\!\!/}
\def\k1S{k_1\!\!\!/}
\def\k2S{k_2\!\!\!/}
\def\qS{q\!\!\!/}
\def\pS{p\!\!\!/}

\def\intm#1#2#3{           {\it Int.J. Mod. Phys. }{\bf #1}, #2 (#3)}

\def\nc#1#2#3{           {\it Nuovo Cim.  }{\bf #1}, #2 (#3)}

\def\pl#1#2#3{           {\it Phys. Lett. }{\bf #1}, #2 (19#3)}

\def\pr#1#2#3{           {\it Phys. Rev. }{\bf #1}, #2 (19#3)}

\def\pro#1#2#3{          {\it Prog. Theor. Phys. }{\bf #1}, #2 (19#3)}

\def\rmp#1#2#3{          {\it Rev. Mod. Phys. }{\bf #1}, #2 (19#3)}

\begin{document}
\begin{titlepage}
\renewcommand{\thefootnote}{\fnsymbol{footnote}}
\begin{center}
\hfill
\vskip 1 cm
{\large \bf Is the axial anomaly really determined in a continuous non-perturbative regularization?}
\vskip 1 cm
{
  {\bf J. L. Jacquot}\footnote{jacquot@lpt1.u-strasbg.fr}
   \vskip 0.3 cm
   {\it Laboratoire de Physique Th\'eorique,
        3 rue de l'Universit\'e,
        67084 Strasbourg Cedex, FRANCE}\\ }
  \vskip 0.3 cm
\end{center}

\vskip .5 in
\begin{abstract}
In the framework of a gauge invariant  continuous and non-perturbative regularization scheme based on the smearing of point like interactions by means of cutoff functions, we show that the axial anomaly, though cutoff independent,  depends on the shape of the cutoff functions.
The standard value for the strength of the axial anomaly is recovered if  we assume that the regularized gauge invariant axial current is in addition local. 
\end{abstract}
\end{titlepage}
\renewcommand{\thepage}{\arabic{page}}
\setcounter{page}{1}
\setcounter{footnote}{0} 
\mysection{Introduction}
 Since its discovery fifty years ago \cite{FUKU}, and its recognition as an intrinsic feature of the regularization of gauge theories in QFT \cite{BARD,JACK1}, the multiple properties of the axial anomaly were extensively studied.
As we know, if a subtracting regularization scheme is used in perturbation theory, like for example the Pauli-Villars regularization \cite{VILLARS1}, the axial anomaly is finite at the one loop order, and its value is determined only once one has  decided which symmetry must be preserved.
This is due to the fact that the integral associated to the triangular Feynman diagram is linear divergent, and hence its finite part becomes ambiguous because it depends of the shift of the loop momentum integration variable \cite{JACK1}.
It is in this sense that Jackiw  \cite{JACK2} emphasized recently that the axial anomaly is an example which shows that radiative corrections can be finite,  but undetermined in QFT.
We show in the framework of QED that the undetermined nature of the axial anomaly gets stronger, if we use a continuous and non-perturbative gauge invariant regularization based of the smearing of point like interactions by means of cutoff functions \cite{JLJ}.
It comes out that at one loop order the form of the axial anomaly is the standard one, but its strength depends on the shape of the cutoff functions if strict locality for the regularized axial current is not assumed.
In order to derive this result we  calculate directly the Green's function $\bra J^{\mu}_5(q) A_{\alpha}(k_1)A_{\beta}(k_2) \ket$ relative to the transition amplitude of the axial current $J^{\mu}_5(q)$ to two photons in momentum space starting from the  regularized equations of motion of QED.
\mysection{The divergence of the axial current} 
First of all we recall briefly the main features of the non-perturbative regularization scheme under consideration \cite{JLJ}.
The regularized action is
\beqa
\label {skthree}
\ S(\psi,\bpsi,A)~&=&~\int\dk ~\bpsi(k) \left(\rho_1^2(k)\kS-m 
\right)\psi(k)~-~e\int d\bar{p} d\bar{p'}~\bpsi(p)A^{\mu}(p-p')\Gamma_{\mu 
}(p,p')\psi(p')\nonumber \\
& &+~\sum_{n=0}^{+\infty} \frac{(ie)^{(n+2)}}{(n+2)!}\int\dk 
d\bar{p} d\bar{p'}~\bpsi(p) K_{n+2}(k,p,p') \psi(p') + S_{Gauge},
\eeqa
where we have used the notation $d\bar{p}$ for $dp/(2\pi)^4$.
The fermionic part of the action is the sum of three terms and is regularized in a gauge invariant manner with the help of the cutoff functions $\rho_i(k) \equiv \rho_i(k^2/ \Lambda^2)$ whose asymptotic forms are
\beq
\label{limcutf} 
\lim_{\Lambda \to \infty}\rho_i (k,\Lambda )~=~1.
\eeq
Apart from the fact that in euclidian space the UV  cutoff functions must be positive and  rapid decreasing  functions of the squared momenta their form is quite arbitrary.
The first term which is the free electron kinetic term,  gives the expression of the regularized free electron propagator.
The second term can be deduced from the standard non-regularized electron photon interaction  if we substitute the bare vertex $\gamma_{\mu}$ by 
\beq
\label {Gamone}
\ \Gamma_{\mu}(p,p')~=~\rho_3(p-p')\left[\rho_2(p)\rho_2(p')\Gamma_{\mu}(p-p') 
+ 
\frac{(p-p')_{\mu}}{(p-p')^2}\left(\rho_1^2(p) \pS-\rho_1^2(p') 
\pS'\right) 
\right].
\eeq
By construction $\Gamma_{\mu}(p,p')$  contains a transverse part relative to the momentum $p-p'$, which is proportional to $\Gamma_{\mu }(p-p')$, where
\beq
\label {Gamtwo}
\Gamma_{\mu }(q)~~=~~\gamma_{\mu }-\qS\frac{q_{\mu }}{q^2}.
\eeq
The third term  is defined by the kernel 
\beqa
\label {skone}
K_{n+2}(k,p,p')&=&F_{n+2}(p-k,k-p')\rho_1^2(k)\kS ~+~i(n+2)  \int \dq~\rho_3(q) A^{\mu}(q)\Gamma_{\mu }(q) \nonumber \\
&& F_{n+1}(p-k,k-q-p') \rho_2(k)\rho_2(k-q) \\
\label {recur}
\  F_{n+1}(p,q)&=&-i\int \dr ~\rho_3(r) \frac{r_{\mu }}{r^2} 
\left(F_n(p-r,q)-F_n(p,q-r)\right)A^{\mu }(r), 
\eeqa
with $F_0(p,q)=((2\pi )^4)^2~\delta (p)\delta (q)$.
This term which describes an infinite set of interactions  between two electrons and any number of photons (at least two), ensures that the vertices of  each n-photons amplitude are automatically constructed with the matrices (\ref{Gamtwo}) and hence are transverse relative to the external photon momenta.
Finally  $S_{Gauge}$ is the sum of the standard non-regularized photon kinetic term, of the gauge fixing terms, and of a new interaction which is quadratic in the photon fields.
The new term which is proportional to the fine structure constant $\alpha$ plays the role of a counterterm for the polarization operator which shows  a quadratic divergence in its transverse part in this regularization scheme.
Moreover this  term which is needed to fix the value of the  photon mass is non-renormalized by higher order  radiative corrections and can be absorbed in the photon propagator \cite{JLJ}.

\begin{figure}[H]
\begin{center}
\epsfig{figure=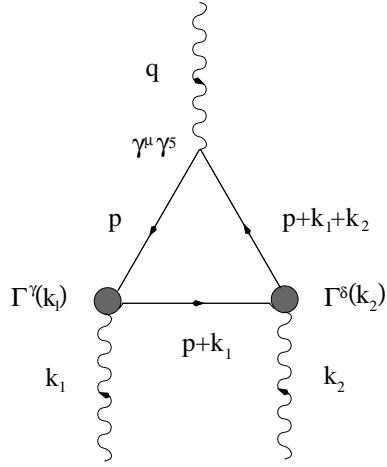,width=5cm}
\caption{The structure of the regularized triangular diagram.}
\label{axialfig1}
\end{center}
\end{figure}

Now we are able to calculate the divergence of the axial current.
From the regularized equation of motion $\bra \delta S(\psi,\bpsi,A)/\delta \bpsi (p) + \eta (p) \ket = 0$ which is deduced from the translational invariance of the regularized partition function of QED in presence of the external  sources  $\eta$, $\bta$ and $J$ for the electron  and photon fields, we  first deduce the vacuum expectation value of the  electron field in presence of the external  sources.
Then, the derivation of this latter expression with respect to the external  sources, allows to  express the  Green's function $\bra J^{\mu}_5(q) A_{\alpha}(k_1)A_{\beta}(k_2) \ket$ as
\beqa
\label {axialone}
 \bra J^{\mu}_5(q)  A_{\alpha}(k_1)A_{\beta}(k_2) \ket &=& e\int d\bar{p} d\bar{p'}~ \bra \bpsi (p-q)\gamma^{\mu }\gamma_5 S(p) A^{\gamma}(p-p')\Gamma_{\gamma}(p,p') \psi(p')A_{\alpha}(k_1)A_{\beta}(k_2) \ket \nonumber \\
& &-~\sum_{n=0}^{+\infty} \frac{(ie)^{(n+2)}}{(n+2)!} \int\dk 
d\bar{p} d\bar{p'}~\bra \bpsi(p-q)\gamma^{\mu }\gamma_5 S(p) K_{n+2}(k,p,p') \nonumber \\
&& \psi (p')A_{\alpha}(k_1)A_{\beta}(k_2) \ket.
\eeqa
In this expression $S(p)$ is the regularized free electron propagator $1/\left(\rho_1^2(p)\pS-m\right)$ and $J^{\mu}_5(q)$ is the Fourier transform of the axial current $\bpsi \gamma^{\mu }\gamma_5 \psi$.
Expressing the connected part of the Green's function (\ref{axialone}) in terms of 1PI functions with the help of Schwinger's sources technique, and keeping only the lower order terms in the coupling constant for the vertex function and for the electron propagator, we get
\beqa
\label {axialtwo}
 \bra J^{\mu}_5(q)  A_{\alpha}(k_1)A_{\beta}(k_2) \ket &=& ie^2 \delta (q+k_1+k_2)\rho_3(k_1)\rho_3(k_2) D_{\alpha \gamma}(k_1)  D_{\delta \beta}(k_2) \int dp ~\rho_2 (p+k_1+k_2) \nonumber \\
 && \rho_2^2 (p+k_1) \rho_2 (p) Tr \gamma^{\mu}\gamma_5 S(p) \Gamma^{\gamma}(k_1) S(p+k_1) \Gamma^{\delta} (k_2) S(p+k_1+k_2) \nonumber \\
&& +~(\alpha,k_1 \longleftrightarrow \beta, k_2 ),
\eeqa
where $ D_{\mu\nu}(p)$ is the photon propagator.
Here we must stress that this result was obtained in two steps.
The first term of (\ref{axialone}),   gives an expression similar to  (\ref{axialtwo}), but with the matrices $\Gamma^{\gamma}(k_1)$ and $\Gamma^{\delta}(k_2)$ replaced respectively by their  non transverse counterparts $\Gamma^{\gamma} (p+k_1,p)$ and $\Gamma^{\delta}(p+k_1,p-q)$ \footnote{This expression converges formally to the standard non regularized triangle amplitude when the cutoff tends to infinity.}.
It is only when we take into account the second term of (\ref{axialone}) that the non transverse part of the vertices associated to the external photons lines cancel algebraically after some judicious shift of integration variable.
We define the 1PI function $\Gamma_5^{\mu \gamma\delta}$ associated to the Green's function (\ref{axialtwo}) as
\beq
\label {axialthree}
 \bra J^{\mu}_5(q)  A_{\alpha}(k_1)A_{\beta}(k_2) \ket ~=~(2\pi)^4 \delta (q+k_1+k_2) D_{\alpha \gamma}(k_1)  D_{\delta \beta}(k_2) \Gamma_5^{\mu \gamma\delta}(k_1,k_2).
\eeq 
In the same way we define also  the 1PI function $\Gamma_5^{ \gamma\delta}$ associated to the amplitude $\bra J_5(q)\\ A_{\alpha}(k_1)A_{\beta}(k_2) \ket$, where $J_5(q)$ is the Fourier transform of the pseudoscalar density $\bpsi\gamma_5 \psi$. 
Owing to the transversality property of the matrices (\ref{Gamtwo}), we see directly by inspection, that the amplitude $\Gamma_5^{\mu \gamma\delta}$ (\ref{axialthree})  is transverse with regard to the external photons lines, i.e.
\beq
\label {trans1}
k_{ 1\gamma}\Gamma_5^{\mu \gamma\delta}~=~k_{2\delta}\Gamma_5^{\mu \gamma\delta}~=~0.
\eeq
Notice that due to the gauge invariant regularization used, the structure of the 1PI function (\ref{axialthree}) which is represented by the triangular diagram of Fig. \ref{axialfig1} can also be deduced in a straightforward manner, if we impose from the beginning the conditions (\ref{trans1}).
\newfloat{figure}{H}{cap}
Is the axial current conserved as well?
If we contract the two members of the relation (\ref{axialtwo}) by the momentum $(k_1+k_2)_{\mu}$ of the two incoming photons, use the identity
\beq
\label {idenpropa1}
\qS~=~\rho_1^{-2}(p+q) S^{-1}(p+q)-\rho_1^{-2}(p) S^{-1}(p) +m \left(\rho_1^{-2}(p+q)-\rho_1^{-2}(p)\right),
\eeq
and shift the integration variable $p$ by  $p-k_1$,  the definition (\ref{axialthree}) leads to the regularized expression   
\beqa
\label {axialfour}
(k_1+k_2)_{\mu}\Gamma_5^{\mu \gamma\delta}(k_1,k_2)&=&i\frac{e^2}{(2\pi)^4} Tr \int dp~ \left[ \rho^2_2(p) \rho_2(p-k_1)\rho_2(p+k_2) \left(\rho_1^{-2}(p+k_2)\gamma_5 S(p-k_1)\right. \right.\nonumber \\
 && \left.\left.\Gamma^{\gamma}(k_1) S(p) \Gamma^{\delta} (k_2) + \rho_1^{-2}(p-k_1)\gamma_5\Gamma^{\gamma}(k_1)S(p) \Gamma^{\delta} (k_2)S(p+k_2) \right)  \right.\nonumber \\ &&+\left.\left. 2m\rho^2_2(p)\rho_2(p-k_1)\rho_2(p+k_2)\rho_1^{-2}(p+k_2)\gamma_5 S(p-k_1)\Gamma^{\gamma}(k_1)\right. \right.\nonumber \\
&& \left.\left.  S(p)\Gamma^{\delta} (k_2)S(p+k_2) \right.\right] +~(\gamma,k_1 \longleftrightarrow \delta, k_2 ).
\eeqa
Due to the properties of the  $\gamma_5$ matrix, and those of the totally antisymmetric tensor $\epsilon^{\mu\nu\gamma\delta}$, only the terms containing the product of the two matrices $\gamma^{\gamma}$ and $\gamma^{\delta}$ contribute to the trace of (\ref{axialfour}).
In this case it is easy to see that the  term proportional to $2m$, is finite and is nothing else but the standard amplitude $\Gamma_5^{ \gamma\delta}$ \cite{POKO}.
Thus we obtain for  (\ref{axialfour})
\beqa
\label {axialfive}
(k_1+k_2)_{\mu}\Gamma_5^{\mu \gamma\delta}(k_1,k_2)&=&8\frac{e^2}{(2\pi)^4} \epsilon^{\gamma\delta\rho\sigma}  \int dp~  \rho^2_2(p) \rho_2(p-k_1) \rho_2(p+k_2) \rho^{-2}_1(p) \rho_1^{-2}(p-k_1) \nonumber \\
&& \rho_1^{-2}(p+k_2)\left[ k_{1\rho}p_{\sigma} D(p)D(p-k_1) +k_{2\rho}p_{\sigma} D(p)D(p+k_2) \right] \nonumber \\
&& +2m\Gamma_5^{ \gamma\delta}(k_1,k_2).
\eeqa
Here we have rewritten the regularized free electron propagator $S(p)$ as $\rho^{-4}_1(p) ( \rho^2_1(p)\pS+m ) D(p)$, where
\beq
\label {propa1}
D(p)~=~\frac{1}{p^2-m^2\rho^{-4}_1(p)}.
\eeq
Now some remarks have to be made.
1) From the property (\ref{limcutf})  we obtain the standard linearly divergent piece of the amplitude $(k_1+k_2)_{\mu}\Gamma_5^{\mu \gamma\delta}$\cite{POKO} which induces the axial anomaly if we formally interchange the limit with the integral symbol.
2) The pseudotensorial structure of the first term of (\ref{axialfive}) implies that the two regularized integrals on the right-hand side of (\ref{axialfive}) do not vanish only because the product of the cutoff functions is a function of both the external momenta $k_1$ and $k_2$.
3) The first term on the right hand-side of (\ref{axialfive}) is independent of $m$ because its derivative with respect to   $m$ is finite and  zero.
4) The first term on the right hand-side of (\ref{axialfive}) vanishes as $k_1=0$ or  $k_2=0$.
Therefore, if we parametrize respectively the shape of the UV  cutoff functions $\rho_1(k) \equiv \rho(a k^2/ \Lambda^2)$ and  $\rho_2(k) \equiv \rho(b k^2/ \Lambda^2)$   by two real numbers  $a$ and $b$, it follows from dimensional analysis and from the requirement of Lorentz invariance that the right-hand side of (\ref{axialfive}) is given by
\beq
\label {axialsix}
(k_1+k_2)_{\mu}\Gamma_5^{\mu \gamma\delta}(k_1,k_2)~=~i\frac{e^2 }{2\pi^2} c(a,b) \epsilon^{\gamma\delta\rho\sigma}   k_{1\rho} k_{2\sigma} +2m\Gamma_5^{ \gamma\delta}(k_1,k_2),
\eeq
where $c(a,b)$ is a dimensionless and finite real function.
Moreover if we rescale the parameters $a$ and $b$ respectively by  $as$ and $bs$, where $s$ is a real constant, it is easy to see from (\ref{axialfive}), (\ref{propa1}) and (\ref{axialsix}) that $c(as,bs)~=~c(a,b)$.
This implies that we can write
\beq
\label{relc1}
c(a,b)~\equiv~c(x),
\eeq
$x$ being the ratio $b/a$. 
The fact that the function $c(a,b)$ is an homogeneous function of zeroth order with respect to the real variables  $a$ and $b$, shows that the strength of the axial anomaly (\ref{axialsix}) which is due to the regularization of an evanescent operator is  sensitive to the relative shape of the cutoff functions, but is independent of the rescaling of the cutoff $\Lambda$.
Next we  show  that this is indeed the case.

We choose for example the following  cutoff functions
\beq
\label {cutoff1}
\rho_1(p)~=~e^{a\frac{p^2}{\Lambda^2}}, ~~~~~\rho_2(p)~=~e^{b\frac{p^2}{\Lambda^2}},
\eeq
whose   shape  is  parametrized  by the real positive numbers  $a$ and $b$.
In addition the signature of the euclidian metric is taken as $(-1. -1.-1.-1)$.
If we suppose that 
\beq
\label {cutoff2} 
\frac{b}{a} > \frac{3}{2}
\eeq
then the function $D(p)$ can be substituted by $1/(p^2-m^2)$ without changing the result of the integration since the product of the cutoff functions is proportional to $  \exp (p^2 (4b-6a)/ \Lambda^2)$ in the integral  on the right hand-side of (\ref{axialfive}).
In this case we can perform an analytical calculation \footnote{ The integrals necessary to the calculation are listed in the appendix of \cite {JLJ}.}, and  we obtain  for the function (\ref{relc1}),
\beq
\label {funcc1}
c(x)~=~\frac{x-2}{4x-6}.
\eeq
The point of discontinuity of the function (\ref{funcc1}) reflects the fact that the integral on the  right hand-side of (\ref{axialfive}) is not regularized in the limit of exact vanishing electron mass when the value of the ratio $x$ is $3/2$.

Since the first piece on the right hand-side of (\ref{axialfive}) is regularized for all positives values of $a$ and $b$ and  since the function  $c(a,b)$ is independent of $m$, it is worth to notice that, even though the result (\ref{funcc1}) was derived under the condition  (\ref{cutoff2}),  this result holds for any positive values of $x$.
It follows from (\ref{funcc1}) that the strength of the axial anomaly is not fixed if the relative shape of the cutoff functions is not constrained.
We  recover the standard one loop value $e^2/(2\pi)^2$ for the coupling constant of the axial anomaly when the function (\ref{funcc1}) verifies 
\beq
\label {equc1}
c(x)~=~1,
\eeq
i.e. when the ratio $x$ of the shapes of the cutoff functions is constrained by the condition $x~=~\frac {4}{3}$.

Is the result (\ref{axialsix}) specific to the regularization used?
Can we understand the physical meaning of the  condition (\ref{equc1})?
In order to answer  these questions we  study the structure of the  regularized axial current.
\mysection{The regularized axial current}
In order to calculate the 1PI function $\Gamma_5^{\mu \gamma\delta}$ (\ref{axialthree}) we start from the expression of the regularized amplitude (\ref{axialtwo}), and  write
\beq
\label {axialseven}
\Gamma_5^{\mu \gamma\delta}(k_1,k_2)~=~I_1^{\mu \gamma\delta}(k_1,k_2) + I_2^{\mu \gamma\delta}(k_1,k_2) +~(\gamma,k_1 \longleftrightarrow \delta, k_2 ).
\eeq
The first term $I_1^{\mu \gamma\delta}$ is  due only to the contributions of the Dirac matrices which enter in  the vertices (\ref{Gamtwo}) and converges formally to the standard non regularized triangle graph when the cutoff tends to infinity.
As for the second piece $I_2^{\mu \gamma\delta}$  it is only a sum of terms proportional to $k_1^{\gamma}/k_1^2$ and  $k_2^{\delta}/k_2^2$.
Due to the properties of the $\gamma_5$ matrix, the term proportional to  $k_1^{\gamma}k_2^{\delta}/k_1^2k_2^2$ does not contribute.
If we use the identity (\ref{idenpropa1}), we obtain   the simple expression for  $I_2^{\mu \gamma\delta}$ in a straightforward manner, i.e.
\beqa
\label {currentaxial1}
I_2^{\mu \gamma\delta}(k_1,k_2)&=&i\frac{e^2}{(2\pi)^4} Tr \int dp~ \frac{k_1^{\gamma}}{k_1^2}\gamma_5 \gamma^{\mu} \left[\rho_2(p+k_1+k_2) \rho_2(p) \left(\rho^{-2}_1(p+k_1)\rho^{2}_2(p+k_1) \right. \right.\nonumber \\
&& \left.\left. - \rho^{2}_2(p+k_2)\rho^{-2}_1(p+k_2) \right) S(p)\gamma^{\delta} S(p+k_1+k_2) + \rho_2(p+k_2) \rho_2(p)\right.\nonumber \\  &&\left. \left(\rho_2(p+k_2) \rho^{-2}_1(p+k_1+k_2)\rho_2(p+k_1+k_2) -\rho_2(p-k_1)\rho_2(p)\rho^{-2}_1(p-k_1) \right) \right.\nonumber \\
&& \left. S(p)\gamma^{\delta}S(p+k_2) \right].
\eeqa
This integral  vanishes formally and  is linear divergent when the cutoff tends to infinity.
The two divergences, which are in fact logarithmtic \footnote{This is because the result of the integration is a linear combination of the external momenta.} cancel each other, and with the particular choice (\ref{cutoff1},\ref{cutoff2}) for the cutoff functions, we obtain the finite expression
\beq
\label {currentaxial2}
I_2^{\mu \gamma\delta}(k_1,k_2)~=~i\frac{e^2}{4\pi^2} \left(1-c(a,b) \right) \frac{k_1^{\gamma}}{k_1^2} k_{1\rho}  k_{2\sigma}\epsilon^{\mu\delta\rho\sigma},
\eeq
with $c(a,b)$ given by (\ref{relc1},\ref{funcc1}).
The calculation of the first term of (\ref{axialseven}) is more involved.
Expanding the trace, and isolating the finite term proportional to $m^2$, we obtain in a intermediate step
\beqa
\label {currentaxial3}
I_1^{\mu \gamma\delta}(k_1,k_2)&=&4\frac{e^2}{(2\pi)^4} \epsilon^{\mu\gamma\delta\rho} \int dp~g(p,k_1,k_2)p_{\rho} +i\frac{e^2}{(2\pi)^4} \int dp~f(p,k_1,k_2) \left[4i\epsilon^{\mu\gamma\delta\rho} (k_1-k_2)_{\rho} p^2 \right.\nonumber \\
&& \left. +8i\epsilon^{\mu\gamma\delta\rho} pk_1k_{2\rho} -8i\epsilon^{\mu\gamma\rho\sigma}p_{\rho}k_{1\sigma}(p+k_2)^{\delta} -8i\epsilon^{\mu\delta\rho\sigma} (p-k_1)^{\gamma}p_{\rho}k_{2\sigma}\right.\nonumber \\
&& \left. -Tr\gamma^5\gamma^{\mu}\gamma^{\gamma} \pS \kS_1 \kS_2 \gamma^{\delta} \right] +4\frac{e^2}{(2\pi)^4} m^2\epsilon^{\mu\gamma\delta\rho}(k_1-k_2)_{\rho} \int dp~f(p,k_1,k_2),
\eeqa
where we have defined the functions $f(p,k_1,k_2)$ and $g(p,k_1,k_2)$ in terms of $D(p)$ (\ref{propa1}) as
\beqa
\label {deffunc1}
f(p,k_1,k_2)&=&\rho_2(p+k_2)\rho_1^{-2}(p+k_2) \rho_2^2(p)\rho_1^{-2}(p)\rho_2(p-k_1)\rho_1^{-2}(p-k_1)   D(p+k_2)D(p)\nonumber \\ 
&&D(p-k_1) \nonumber \\ 
g(p,k_1,k_2)&=&f(p,k_1,k_2)D(p)^{-1}.
\eeqa
If it is not regularized, the remaining part of $I_1^{\mu \gamma\delta}$  contains two kinds of UV divergences.
The first divergence is linear and comes from the term containing the function $g(p,k_1,k_2)$.
This divergence is reduced to the sum of a logarithmtic one, which  is proportional to the asymptotic form $(k_1-k_2)_{\rho}\epsilon^{\mu\gamma\delta\rho} \int dp~/p^4$,  and of a constant which depends on the shape of the cutoff functions. 
The second divergence which comes from the expression in the brackets is purely logarithmtic, and  is induced by the asymptotic forms  $\epsilon^{\mu\gamma\delta\rho}(k_1-k_2)_{\rho} \int dp ~p^2/p^6$ and $k_{1\sigma}\epsilon^{\mu\gamma\rho\sigma} \int dp~ p_{\rho}p^{\delta}/p^6 + k_{2\sigma}\epsilon^{\mu\delta\rho\sigma} \int dp~ p_{\rho}p^{\gamma}/p^6$.
Because  the two kind of divergences cancel each other in the regularized form (\ref{currentaxial3}), it follows that the term $I_1^{\mu \gamma\delta}$ is indeed finite.
Finally when the cutoff functions (\ref{cutoff1}) obey the condition (\ref{cutoff2}) we get the analytical expression for the complete amplitude $\Gamma_5^{\mu \gamma\delta}$ (\ref{axialseven}) 
\beqa
\label {currentaxial4}
\Gamma_5^{\mu \gamma\delta}(k_1,k_2)&=&i\frac{e^2}{2\pi^2} \left[ k_{1\rho} \epsilon^{\mu\gamma\delta\rho} A_1(k_1,k_2) +  k_{2\rho}\epsilon^{\mu\gamma\delta\rho} A_2(k_1,k_2)  +  k_{1\rho}k_{2\sigma} k_1^{\delta}\epsilon^{\mu\gamma\rho\sigma} A_3(k_1,k_2) \right. \nonumber \\ 
&& \left. +k_{1\rho}k_{2\sigma} k_2^{\delta}\epsilon^{\mu\gamma\rho\sigma}A_4(k_1,k_2) + k_{1\rho}k_{2\sigma} k_1^{\gamma}\epsilon^{\mu\delta\rho\sigma} A_5(k_1,k_2) \right. \nonumber \\
&& \left. + k_{1\rho}k_{2\sigma} k_2^{\gamma} \epsilon^{\mu\delta\rho\sigma} A_6(k_1,k_2) \right]
 -i\frac{e^2}{2\pi^2} m^2 (k_1-k_2)_{\rho}\epsilon^{\mu\gamma\delta\rho} I_{00}(k_1,k_2).
\eeqa
In this expression all the $A_i$ are finite, and are given in terms of the integrals \footnote{Our definition of the integrals  $I_{st}$ is similar to that of Rosenberg's \cite{ROSEN1} if $k_1$ is exchanged with $k_2$.}
\beq
\label {introsen1}
I_{st}(k_1,k_2)~=~\int_0^1 dx \int_0^{1-x} dy \frac{x^sy^s}{-x(1-x)k_1^2 -y(1-y)k_2^2 -2xyk_1k_2+m^2},
\eeq
by
\beqa
\label {currentaxial5}
A_1(k_1,k_2)&=&\frac{1}{2} \left( c(a,b)+ k_1^2 I_{10}(k_1,k_2) -k_2^2 I_{01}(k_1,k_2) \right) \\
\label {currentaxial6}
A_2(k_1,k_2)&=&A_1(k_1,k_2) -c(a,b)\\
\label {currentaxial7}
A_3(k_1,k_2)&=&-2I_{11}(k_1,k_2)~~~~~~A_6(k_1,k_2)~=~-A_3(k_1,k_2)\\
\label {currentaxial8}
A_4(k_1,k_2)&=&-2\left(I_{01}(k_1,k_2)-I_{02}(k_1,k_2) \right) -\frac{1}{2k_2^2} \left(1-c(a,b) \right)
 \\
\label {currentaxial9}
A_5(k_1,k_2)&=&2\left(I_{10}(k_1,k_2)-I_{20}(k_1,k_2) \right) +\frac{1}{2k_1^2} \left(1-c(a,b) \right),
\eeqa
with $c(a,b)$ given in this case by (\ref{relc1}) and (\ref{funcc1}).
Since the integrals $I_{st}$ verify the property
\beq
\label {introsen2}
k_2^2\left( I_{01}(k_1,k_2) -2I_{02}(k_1,k_2) \right)~=~k_1^2\left( I_{10}(k_1,k_2) -2I_{20}(k_1,k_2) \right)
\eeq
the amplitude $\Gamma_5^{\mu \gamma\delta}$ is transverse with regard to the externals photons lines, and as expected we recover  the expression (\ref{axialsix}) for the divergence of the axial current.
 
Now we show that the expression of the amplitude (\ref{currentaxial4}) is quite general, in the sense that its form is unchanged when   the  value of the factor $c(a,b)$ is not constrained by the condition (\ref{cutoff2}).

From the property $g(-p,k_1,k_2)=g(p,k_2,k_1)$, it follows that the finite cutoff dependent part of the integral
$\epsilon^{\mu\gamma\delta\rho} \int dp~g(p,k_1,k_2)p_{\rho}$ is proportional to $\epsilon^{\mu\gamma\delta\rho}(k_1-k_2)_{\rho}$.
Then, if we compare the divergence of the axial current obtained from (\ref{currentaxial4}) with the general expression (\ref{axialsix}), we conclude that the factor $c(a,b)$ which enters in $A_1$ (\ref{currentaxial5}) and $A_2$ (\ref{currentaxial6}) is just the factor $c(a,b)$ entering in (\ref{axialsix}).
Finally the requirement of gauge invariance implies that the factor $c(a,b)$ which enters in $A_4$ (\ref{currentaxial8}) and $A_5$ (\ref{currentaxial9}), is again the general factor $c(a,b)$ of (\ref{axialsix}).
We can notice that the factors $A_4$ and $A_5$, are similar to the finite factors obtained by Rosenberg \cite{ROSEN1} with dimensional arguments, if and only if the value of the factor $c(a,b)$ is one (\ref{equc1}).
In this case, the axial anomaly (\ref{axialsix})  has the standard numerical value and the contribution (\ref{currentaxial2}) of $I_2^{\mu \gamma\delta}$ vanishes.
Since the integral $I_2^{\mu \gamma\delta}$ vanishes formally as the cutoff goes to infinity and is thus due to the contribution of an evanescent operator,  the fact that the axial anomaly can be undetermined is directly related to the non uniform convergence of the regularized integrals, as we will now see.
\mysection{Discussion}
In perturbation theory the amplitudes are in general regularized through the regularization of individual Feynman diagrams.
If we define the standard non regularized amplitude $\Gamma_5^{\mu \gamma\delta}(k_1,k_2) \equiv \int dp~ \mathcal{A}_5^{\mu \gamma\delta}(p,k_1,k_2)$, we know from  \cite{JACK1} that the difference of two non regularized triangle graphs, which differ only from a shift of integration variable, is given by
\beq
\label {anoJack1}
\int dp~  \mathcal{A}_5^{\mu \gamma\delta}(p,k_1,k_2) -\int dp~  \mathcal{A}_5^{\mu \gamma\delta}(p+ak_1+bk_2,k_1,k_2) \propto (b-a) \epsilon^{\mu\gamma\delta\rho}(k_1-k_2)_{\rho}.
\eeq
The term on the right-hand side  arises from a surface term when the remaining integral is evaluated symmetrically.
This term which induced the anomaly is undetermined because it contains  the arbitrary real constants $a$ and $b$.
Thus if we regularize the amplitude $\Gamma_5^{\mu \gamma\delta}$ by subtracting the divergence \cite{ADLER1} from its integrand, the undetermined contribution to the finite part reflected by (\ref{anoJack1}), is suppressed if we impose a symmetry condition like for instance gauge invariance.
Since, due to dimensional arguments the $A_i$ for $i \geq 3$ are finite \cite{ROSEN1}, this method is equivalent to fix uniquely, by the requirement of gauge invariance, the finite part of $A_1$ and $A_2$ \cite{ROSEN1}  without any kind of regularization.
The net result is that one obtains the standard numerical value for the axial anomaly.
As we have seen, the conclusion is different when we start with regularized amplitudes by means of cutoff functions.

The reason is the following.
Suppose that we regularize the standard triangle graph by hand by  substituting the free electron propagator $S(p)$ by $S(p)\rho(p)$, where $\rho(p)$ is a cutoff function of the kind (\ref{cutoff1}) \footnote{Such a method was used in order to regularize  the  divergence of the axial current in \cite{ZINN1} in a non gauge invariant manner.}.
Then the regularized amplitude 
\beq
\label {reghand}
 \Gamma_{5Reg}^{\mu \gamma\delta}(k_1,k_2) \equiv  \int dp~ \mathcal{A}_5^{\mu \gamma\delta}(p,k_1,k_2,\Lambda)
\eeq
is finite, non gauge invariant and  given by (\ref{currentaxial4}) if we suppress respectively the term proportional to $1/k_2^2$ and $1/k_1^2$ in (\ref{currentaxial8}) and (\ref{currentaxial9}).
It follows that the regularized form of  the left-hand side of (\ref{anoJack1}) is now zero because  the two integrals are invariant under a change of variable.
Since $\lim_{\Lambda \to \infty} \mathcal{A}_5^{\mu \gamma\delta}(p,k_1,k_2,\Lambda) =\mathcal{A}_5^{\mu \gamma\delta}(p,k_1,k_2)$, the discrepancy between the two results for the left hand-side  of (\ref{anoJack1}) is a consequence of the non uniform convergence of the regularized integrals.
In this case the fact that the triangle graph is undetermined is not due to the shift of integration variable, but to the arbitrary choice which we can make for the cutoff functions.
Is the strength of the axial anomaly fixed by gauge invariance in a regularization based on the introduction of cutoff functions in momentum space?

The structure of the 1PI function $\Gamma_5^{\mu \gamma\delta}$  (\ref{currentaxial4}) associated to the regularized triangular diagram is general and is independent of the regularization scheme.
The conditions which are necessary for the gauge invariance of the regularized amplitude (\ref{currentaxial4}) are \cite{ROSEN1}
\beqa
\label {condgauge1}
A_1(k_1,k_2)&=&k_1k_2A_3(k_1,k_2)+k_2^2A_4(k_1,k_2) \nonumber \\
A_2(k_1,k_2)&=&k_1k_2A_6(k_1,k_2)+k_1^2A_5(k_1,k_2).
\eeqa
It follows that there are two possibilities in order to regularize the triangular diagram in a gauge invariant manner.

1) Only the factors $A_1$ and  $A_2$ which enter  the conditions  (\ref{condgauge1}) are functions of the parameters which define the shapes of the cutoff functions.
This is the case in perturbation theory, where  each diagram is regularized independently by multiplying the free propagators or the free vertices by suited cutoff functions.
Then the  conditions (\ref{condgauge1}) of gauge invariance impose that these parameters must verify a relation similar to (\ref{equc1}).
If the overall cutoff function which regularized the analogue of the integrals (\ref{currentaxial3},\ref{propa1}) depends on more than one parameter, then the equation (\ref{equc1}) can have a solution.
The strength of the axial anomaly is then fixed to its standard value by the requirement of gauge invariance.
In this case it is worth to notice that  in general only the triangle graph is regularized in a gauge invariant manner, but not the other diagrams. 
2) By adding an evanescent operator we allow the factors $A_3$, $A_6$, or the factors $A_4$,  $A_5$ to depend also on the parameters of the cutoff functions.
In this case the requirement of gauge invariance  (\ref{condgauge1}) fixes the structure of the former factors in terms of the parameters of the cutoff functions, but does not imply any supplementary constraints for these parameters.
As a result the strength of the axial anomaly stays undetermined.

It is just the second   possibility which is realized when we use a gauge invariant non-perturbative regularization scheme  based on the smearing of the point like interaction by the introduction of cutoff functions.
In this case the relations (\ref{condgauge1}) are fulfilled at the onset without any supplementary constraints for the parameters of the cutoff functions, because now $A_4$ (\ref{currentaxial8})  and  $A_5$  (\ref{currentaxial9}) contain a term which is proportional to $ \left(1-c(a,b) \right)$.
These terms are induced by the regularization which is used and arise from the contributions (\ref{currentaxial2}) to the regularized amplitude  $ \Gamma_5^{\mu \gamma\delta}$ (\ref{axialseven}) of the integrals (\ref{currentaxial1}) which vanish formally when the cutoff tends to infinity.
The contributions (\ref{currentaxial2}) of the evanescent operators (\ref{currentaxial1}) to the regularized axial current defined by (\ref{axialthree}) can be expressed in operator language in terms of the dual $F^{*\mu \nu} \equiv \epsilon ^{\mu \nu \rho \sigma} F_{\rho \sigma} /2$ as
\beq
\label {evanes1}      
I_2^{\mu}~=~- \frac{e^2}{4\pi ^2}\left(1-c(a,b) \right) \partial_{\nu} (F^{*\mu \nu} \frac{\partial_{\alpha}}{\partial^2}A^{\alpha}).
\eeq
This current is conserved, explicitly gauge dependent and non local.
In addition to gauge invariance, if we impose that  the regularized axial current $J^{\mu}_5$ must be  local, the function (\ref{relc1}) must verify the constraint  (\ref{equc1}).
Therefore the strength of the axial anomaly is fixed uniquely to its standard value if the regularized axial current is gauge invariant and strictly local.

Can we understand the result (\ref{axialsix},\ref{currentaxial4}) by comparison with other gauge invariant regularization scheme?

If we exclude dimensional regularization which is in some sense too formal and not adapted to Feynman diagrams containing the $\gamma_5$ matrix, we first study the connection with Pauli-Villars regularization.
In this scheme the regularized fermions loops are obtained by integration over massive regulator fields of negative norm.
Then, as we know, the axial anomaly is given by \cite{POKO}
\beqa
\label {pauli1}
(k_1+k_2)_{\mu}\Gamma_{5Reg}^{\mu \gamma\delta}(k_1,k_2)&=&\lim_{m \to 0} \lim_{M \to \infty} (k_1+k_2)_{\mu}\left (  \Gamma_{5}^{\mu \gamma\delta}(m,k_1,k_2) -\Gamma_{5}^{\mu \gamma\delta}(M,k_1,k_2) \right )\nonumber \\
&=&\lim_{m \to 0} \lim_{M \to \infty} \left ( 2m \Gamma_{5}^{ \gamma\delta}(m,k_1,k_2) -2M \Gamma_{5}^{\gamma\delta}(M,k_1,k_2) \right ) \nonumber \\
&=&-i\frac{e^2 }{2\pi^2} \epsilon^{\gamma\delta\rho\sigma} k_{1\rho} k_{2\sigma} .
\eeqa
We obtain the same result if we replace the non-regularized amplitudes $(k_1+k_2)_{\mu}\Gamma_{5}^{\mu \gamma\delta}$ on the right-hand
 side of (\ref{pauli1}) by their regularized form (\ref{axialsix}).
In this case the standard value of the axial anomaly is just given by the difference between the amplitudes (\ref{axialsix})  of the triangle graph relative to a massless electron and that of an electron with  infinite mass.

In lattice regularization  the known result for the axial anomaly of QED is recovered if the regularized gauge invariant action converges formally to the standard one when the lattice spacing tends to zero and  locality is assumed \cite{ROTHE1}.
In fact, under these general conditions the  lattice chiral Ward-Takahashi identity can be Taylor subtracted at zero momentum, and then in the continuous limit the Taylor subtracted lattice Feynman integrals which enter this identity are finite.
This implies that the continuous limit of the Taylor subtracted lattice chiral Ward-Takahashi identity converges uniformly to the standard anomalous  Ward-Takahashi identity.
Notice that like in the scheme of Pauli-Villars, the regularization of the continuous limit of the lattice regularization of the chiral Ward-Takahashi identity is performed by subtracting  the potentially divergent terms in the integrand of the lattice Feynman integrals.
\mysection{Conclusion and summary}
We have shown that in a non-perturbative gauge invariant regularization scheme of QED, where the divergences are regularized through the smearing of the point like interactions, the divergence of the axial current can be deduced from the regularized equations of motions.
It follows that at the one loop level, the anomalous Ward-Takahashi identity thus obtained is finite but depends on the shape of the cutoff functions.
Since the latter relation is finite one is not free to make any subtraction, if one assumes that the relevant coupling constants which describe QED are only the  charge and electron mass.
This in turn implies  that if a continuous  non-perturbative  regularization is only restricted to preserve  gauge invariance, the axial anomaly is in general sensitive to the form of the cutoff functions.
Like in lattice regularization the strength of the axial anomaly is fixed uniquely to its standard value if in addition to gauge invariance strict locality is assumed  for the regularized axial current.
In this case the parameters which define the shapes of the cutoff functions are constrained to verify the strong relation (\ref{equc1}).
For instance, if all the  cutoff functions which enter in the regularized action of QED are identical, the standard value for the axial anomaly cannot be recovered.
This feature is not seen in  physical perturbative regularization schemes, even in the  continuous limit of lattice regularization, which is also perturbative, because in all these methods the UV divergences are subtracted off.

In conclusion the axial anomaly in QED is only determined  in a non-perturbative regularization scheme if both gauge invariance and locality of the free axial current are preserved.
If locality is not assumed for this current, the  strength of the axial anomaly is finite and does not depend on the rescaling of the cutoff, but depends on the shape  of the cutoff functions.
\vskip 5mm
\centerline{\bf Acknowledgements}
The author would like to thank J. Polonyi for very useful 
discussions.

\end{document}